\documentclass{article}
\usepackage{color,graphicx}
\usepackage[latin1]{inputenc}
\usepackage[authoryear]{natbib}
\usepackage{amsmath,amsfonts,amssymb,amsthm}
\usepackage{fontenc,url,path}
\usepackage{lscape,epsfig}
\usepackage{hyperref}
\usepackage{fancyvrb} % centering verbatim enviroment
\hypersetup{colorlinks,
            urlcolor=blue,
            citecolor=blue}
\usepackage{booktabs} 
\usepackage{natbib}
\usepackage{dsfont} % E of expected value
\newcommand{\keywordname}{Key Words}
\newcommand{\ackname}{Acknowledgements}

\newcommand{\keywords}[1]{%
    \begin{quote}
    \small\textbf{\keywordname: }{#1}
    \end{quote}
}
   {}{}

\definecolor{dgreen}{rgb}{0.,0.6,0.}
\begin{document}
	
\graphicspath{{Plots/}}

\title{\Large\bf Bayesian support for Evolution: detecting phylogenetic signal in a subset of the primate family}

\author{Patricio Maturana R.${}^1$\thanks{Corresponding author. Address for					correspondence: Department of Statistics, University of Auckland, Private Bag 92019, Auckland 1142, New Zealand. Email address: \href{mailto:p.russel@auckland.ac.nz}{p.russel@auckland.ac.nz} (Patricio Maturana R.)}
%	\qquad Brendon J. Brewer${}^1$ \qquad Steffen Klaere${}^{12}$ \\[1.2ex]
%	{\small ${}^1$Department of Statistics, University of Auckland, New Zealand} \\
%	{\small ${}^2$School of Biological Sciences, University of Auckland, New Zealand}
}

\date{\small\today}
\maketitle

%----------------------------------------------------------------------------------------------------------------------
\begin{abstract}
 
The theory of evolution states that the diversity of species can be explained by descent with modification.  Therefore, all living beings are related through a common ancestor.  This evolutionary process must have left traces in our molecular composition.  In this work, we present a randomization procedure in order to determine if a group of 5 species of the primate family, namely, macaque, guereza, orangutan, chimpanzee and human, has retained these traces in its molecules.  Firstly, we present the randomization methodology through two toy examples, which allow to understand its logic.  We then carry out a DNA data analysis to assess if the group of primates contains phylogenetic information which links them in a joint evolutionary history.  This is carried out by monitoring a Bayesian measure, called marginal likelihood, which we estimate by using nested sampling.  We found that it would be unusual to get the relationship observed in the data among these primate species if they had not shared a common ancestor.  The results are in total agreement with the theory of evolution.


\end{abstract}

\section{Introduction}
\label{Intro}
The theory of evolution states that the diversity of species can be explained by descendants with modification.  \cite{darwin1859origin} was able to provide evidence in favour of his theory, despite the limitations at that time.  Nowadays, technology is a powerful tool which allows to generate a huge quantity  of evidence in favour of this theory.  The support comes from different areas, for instance, Molecular Biology, Paleontology, Biogeography, Biochemistry, Phylogenetics.  The present article is located in the latter which is the study of the evolutionary relationship among groups of organisms based typically on molecular sequencing data.

As in any other field, in phylogenetics data analysis is performed mainly under two statistical approaches: Frequentist and Bayesian.  The latter has gained ground in phylogenetics due to its flexibility to deal with large dataset with complex evolutionary models.  Studying a particular Bayesian measure, the probability that the data have been generated from a treelike evolutionary model, we asses whether the patterns of evolution in the molecular sequencing data (DNA) could reasonably arise due to chance.  In other words, if the theory of evolution was right, the sequence alignments should contain information which connects the species from where the DNA was taken. If it is so, we asses if these patterns can be due to chance acting alone.

To evaluate if these patterns emerged from the molecular data is due to chance, we use a method known as \textit{randomization}.  This method allows to detect if the data contain nonrandom information that links the species in a common evolutionary history.  It performs by comparing a statistic obtained from the data to the distribution of the same statistic obtained from a set of functional data, generated randomly from the original one, which consequently does not contain any phylogenetic signal.  If the data support evolution, their information should be significant enough to be differentiated for that one obtained just by chance.  This technique was already proposed by \cite{Archie:1989} in a nonparametric framework.

This article aims to show in a practical way how the evolution theory is supported for a logical method as it is randomization by studying a Bayesian quantity: the marginal likelihood.  First two toy examples are presented as means to understand the method and then an application on a real dataset which contains part of the primate family is given in order to detect phylogenetic signal.  The description of the statistical methods and phylogenetic models are omitted but the respective references are given.

%%%%%%%%%%%%%%%%%
%%% Section 2 %%%
%%%%%%%%%%%%%%%%%

\section{Randomization}

Randomization is a method used to assess the effect of certain factor or treatment on a variable of interest.  This is carried out by studying the properties of the distribution of a statistic calculated from randomized datasets.  Each of these functional datasets is generated by randomly assigning the observations to the factor/treatment, i.e., the experimental units are relabeled.  The new data will not show any effect of the factor on the variable.  The factor is obviated and any difference between its levels is caused by chance.  This is analogous to shuffling playing cards to eliminate any kind of intervention.

The method compares the statistic of the original data with the distribution of the same statistic of the randomized data.  Such statistic for example can be mean, median, mode or variance.  This method does not need to make any assumption about the population, it just works with the data to make inferences.  Assumptions such as normality or equal variances.   The following example helps to understand the method.  

\subsection{Toy examples}
Consider that we have the marks of a test for 10 students differenced by the method of study (A or B) to which the students were randomly allocated.  The marks are presented in percentage and are shown in Table~\ref{toy_example_data}.  The objective is to determine which of the methods of study is more effective.  Both examples are developed at the same context but they will differ in the dataset.  They could have been treated analytically, but to illustrate randomization in a general way we have used simulations.  They just have didactic purposes and clear patterns have been arbitrarily assigned.

\begin{table}
	\centering	
	\begin{tabular}{cc p{0.5cm} cc}
		\toprule
		%\hline\noalign{\smallskip}
		\multicolumn{2}{c}{Example 1} & & \multicolumn{2}{c}{Example 2} \\	
		\cline{1-2} \cline{4-5}
		Method A & Method B & & Method A & Method B \\
		92.5   &   55.2 & & 18.49 &	94.35\\
		99.8   &   32.0	& &	70.24 & 12.92\\
		75.8   &   49.6	& &	57.33 & 83.34\\
		82.4   &   68.3	& &	16.81 & 46.80\\
		93.2   &   69.3	& &	94.38 & 55.00\\	
		\bottomrule
	\end{tabular}
	\caption{Data for the toy examples}
	\label{toy_example_data}
\end{table}

\begin{table}
	\centering	
	\begin{tabular}{lccc}
		\toprule
		& Method A 	& Method B 	& Difference \\
		\midrule
		Example 1 	& 88.74		& 54.88		& 33.86\\
		Example 2 	& 51.45 	& 58.48  	& -7.03 \\
		\bottomrule
	\end{tabular}
	\caption{Means for each method according to the example.  ``Difference'' depicts the subtraction between the means of Method A and B.}
	\label{tab:means}
\end{table}

% Example 1 	& 88.74	(9.5)	& 54.88	(15.3)	& 33.86\\
% Example 2 	& 51.45 (33.6)	& 58.48 (32.1) 	& -7.03 \\

\subsubsection{Example 1}

\begin{figure}
	\centering
	\includegraphics[scale = 0.4]{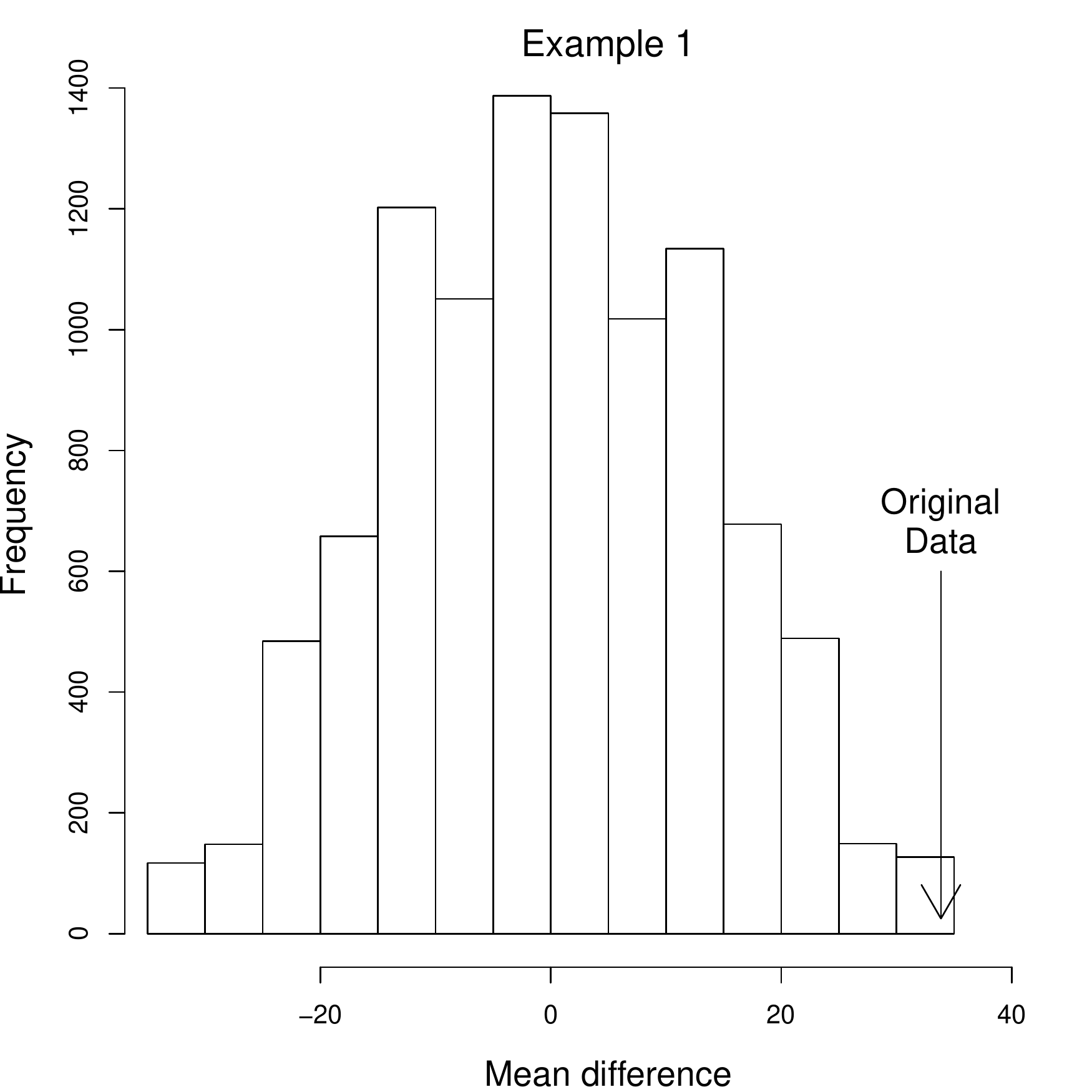}
	\includegraphics[scale = 0.4]{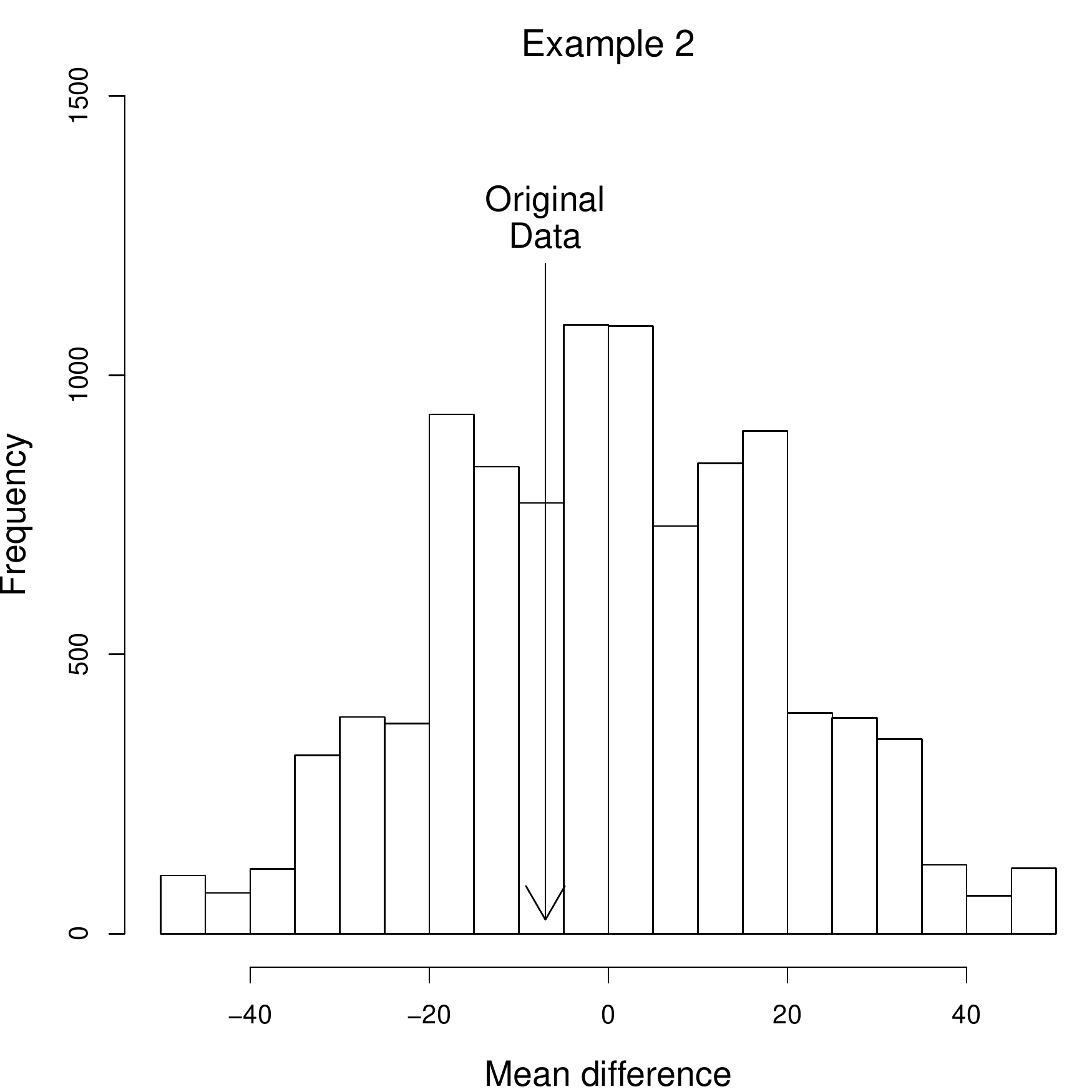}
	\caption{Distribution of the mean of the randomized datasets for the toy examples.  In Example~1, the observed difference is unlikely to have happened under chance acting alone.  On the other hand, in Example~2, this difference could have been just due to chance and nothing to do with which method was used.}
	\label{fig:rand_ex}
\end{figure}

Clearly, method A presents higher marks than method B (see Table~\ref{toy_example_data}, Example~1).  This can be also noticed comparing their means (see Table~\ref{tab:means}).  Apparently method A is better than B.  But can this be due to chance acting alone?  Randomization can give us an idea. \\ 
We generate a new dataset where each mark is assigned randomly to either method A or B.  The number of marks per method is set to 5, as in the original dataset.  Then the difference between the means is calculated and registered.  This procedure is repeated 10,000 times.  The mean differences are plotted in Figure~\ref{fig:rand_ex}. \\  
We can see that the mean difference is around zero.  This is expected because the difference in means is just due to chance.  The effect of the method has been obliterated.  The observed difference, that was calculated from the original data, is 33.86 and located in the right extreme of the distribution.  In case that chance is acting alone, it would be unusual to get an observed difference as big as that observed in the data.  Assuming a well designed experiment, we conclude that method A effectively yields better results than B on average.

\subsubsection{Example 2}

Now consider the data given in Table~\ref{toy_example_data} for Example~2.  In this case both methods yields apparently similar results. The difference between their means is just 7.03 (see Table~\ref{tab:means}, Example 2). But again, can this be due to chance acting alone?  To give an answer we repeat the procedure in Example 1.  The results are shown in Figure~\ref{fig:rand_ex}.

The distribution of the differences between the means of method A and B for the randomized datasets has its center around zero and is relatively symmetric.  Similar characteristics were found in Example 1 because the potential effect of the method of study has been wiped out in both examples.  The observed difference -7.03 is near its center.  When chance is acting alone, this difference is highly probable, unlike Example 1, where the difference in means was unusual under chance acting alone.  Thus, assuming a well designed experiment, we could claim that the methods of study yield similar results, on average, and the observed difference is just due to chance acting alone.  \\

In these cases, we compared the effect of the method of study on the mark mean, but we could have studied any other characteristic, for instance, standard deviation, median, or a specific probability.  In strict rigor, the comparison should be carried out by using an appropriate statistical test, for instance, a $t$-test.  In the next case we will study the probability of the data given the model in order to detect phylogenetic signal in a molecular dataset of 5 primates.

%%%%%%%%%%%%%%%%%%%%%%%%%%%
%%% Primate application %%%
%%%%%%%%%%%%%%%%%%%%%%%%%%%

\section{Phylogenetic analysis}

\begin{figure}
	\centering	
	\includegraphics[scale = 0.40]{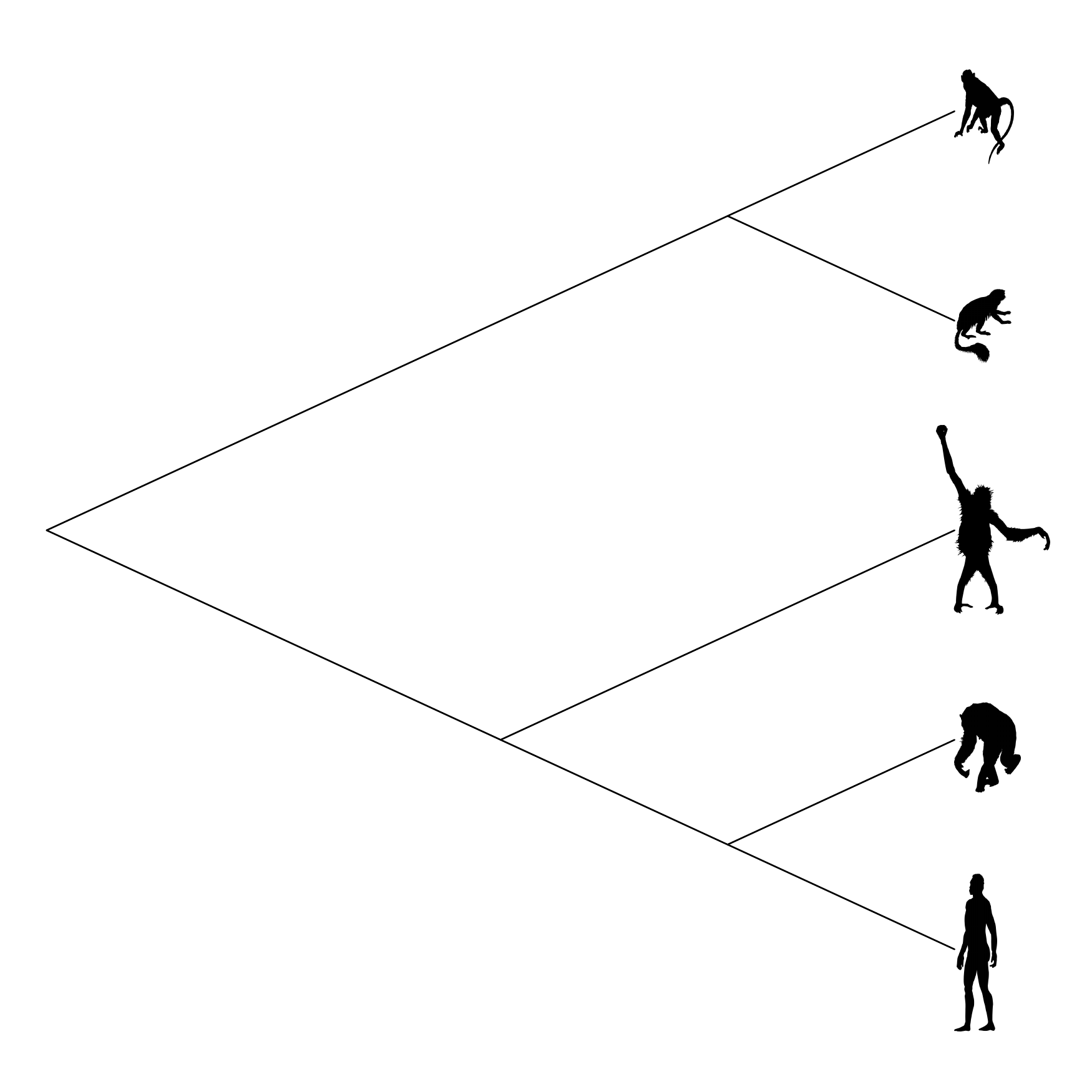}
	\caption{Evolutionary relationship among 5 members of the primate family.  From the top: macaque, guereza, orangutan, chimpanzee and human.  These species are related via common ancestry.}
	\label{tree}
\end{figure}

Now we apply the same concept in order to analyze if a molecular dataset of a group of primates has information about their common evolutionary history.   This is a subset of a dataset which has been previously analysed in the literature \citep{Roos2011}.  This subset contains 5 kinds of primates: macaque, guereza, orangutan, chimpanzee and human.  The alignment corresponds to mitochondrial DNA which has length of 15,727 sites. To wit, the DNA is composed by 4 nucleobases: adenine (\texttt{A}), cytosine (\texttt{C}), guanine (\texttt{G}) and thymine (\texttt{T}).  An extract of the data is shown in Figure~\ref{DNA}.  The relationship among these species is uncontroversial and can be visualized as the tree shown in Figure~\ref{tree}.  Human and chimpanzee share a more recent common ancestor.  This makes them more closely related.  Orangutan is also part of this clade, but with a farer ancestor.  Macaque and guereza form another clade.  All the species are connected through their most recent common ancestor, which is located in the root of the tree (left vertex of the tree). \\
\begin{figure}
	%\centering
	\label{DNA}
\begin{BVerbatim}		
           		1  2  3  4  5  6  7  8  9 10 11 12 13 14 15 16 17 
		Human      C  C  T  A  A  A  A  C  C  C  G  C  C  A  C  A  T  
		Chimpanzee C  T  T  A  A  A  A  C  C  C  T  C  C  A  C  T  T  
		Orangutan  C  C  T  A  A  A  A  C  C  C  T  C  C  A  C  A  T  
		Guereza    C  T  C  A  A  A  A  C  C  C  G  C  A  A  C  C  T  
		Macaque    C  T  T  G  A  A  A  C  C  C  T  C  A  A  C  A  T  
\end{BVerbatim}	
	\caption{Extract of the mitochondrial DNA for 5 species of primates.  Each column represents a site.}	
\end{figure}

In order to eliminate any kind of correlation in the dataset, we permute each site generating a new dataset.  In other words, each site is reordered randomly. For instance, site 2 = \texttt{(C,T,C,T,T)} displayed in Figure~\ref{DNA}, can be permuted as \texttt{(T,C,T,T,C)}.  The theory of evolution \citep{darwin1859origin} states that all organisms are related through common ancestors.  So, if the data were generated by a tree, they should contain this information, unlike in case the data are randomized. 

In the previous examples the mean difference was studied, but now we will study the probability of the data given the model, which will be referred to from now as \textit{marginal likelihood}.  Phylogenetic deals with very small probability values, so it is convenient to work with log values.  The evolutionary relationship among the species is modeled by the tree, which is displayed in Figure~\ref{tree}.  This tree represents the factor to be tested in this analysis, similar to the method of study that was tested in the previous example.  We describe the evolutionary process along the tree assuming a GTR+$\Gamma_4$ model, which is the most general time reversible model.  A good readable material about these models is given in \cite{Yang:2014}.  The prior distributions on the parameters involved in the model are defined in Appendix~1. %~\ref{app1}. 

The calculation of the  marginal likelihood is a challenging problem in phylogenetics, even in simple models.  Therefore, it requires a numerical approximation.  Here we estimate it via Nesting Sampling \citep{Skilling:2006}, algorithm introduced to phylogenetics by \cite{NS_phylo}.  Details of the estimation process are given in Appendix~2. %\ref{app2}.  

\begin{figure}
	\centering
	\includegraphics[scale = 0.32]{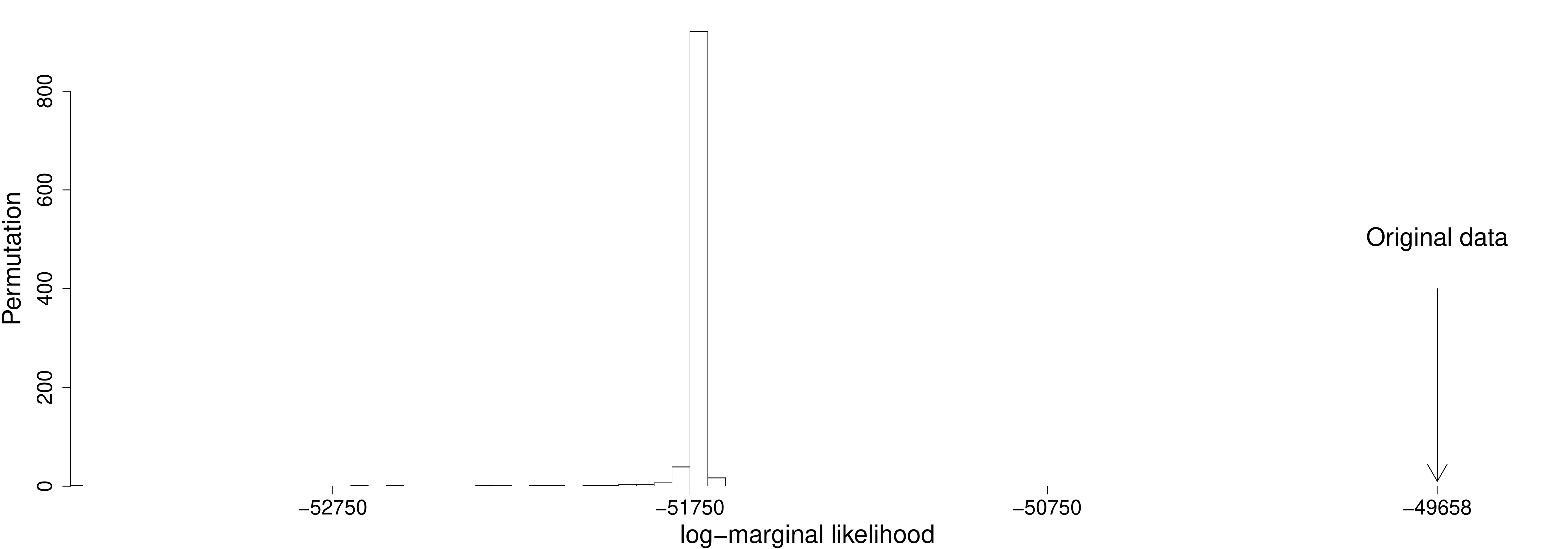}
	\caption{Log-marginal likelihood of the observed data compared to the distribution of this quantity obtained from randomized datasets.  The observed log-marginal likelihood is much higher than that one would expect under chance acting alone.  The information contained in the molecular data of this primate family is more highly probable of being obtained due to common ancestry than just due to chance.}
	\label{perm_ape}
\end{figure}

We generate 1000 randomized datasets and calculate, for each one, their log-marginal likelihoods.  Also, we estimate this quantity for the original dataset.  The results are shown in Figure~\ref{perm_ape} and the descriptive statistics in Table~\ref{descriptive_stats}.

The estimates for the randomized data fluctuate between -53484 and -51675 with a mean of -51737.  On the other hand, the observed log-marginal likelihood estimate is -49658 (with a standard deviation of 0.73).  This is located at the right side of the distribution of the log-marginal likelihoods for the randomized datasets, approximately 26 standard deviation away from the mean. 

Following the reasoning of Example 1, we conclude that it would be unusual that an observed log-marginal likelihood would be as large as the one observed in the data when chance is acting alone.  The probability that the original data has been generated by the tree structure is much higher than the randomized datasets have.  This means that the patterns in the DNA are more likely to be explained by the treelike structure than just to occur due to chance.  In other words, the data contain phylogenetic information that cannot be explained only by chance.  The mitochondrial DNA has retained the common evolutionary history of these species, and our analysis has showed that it would have been highly unlikely to obtain this disposition of the bases in the data as a result of pure chance.  This is evidence which supports the tree structure behind the evolutionary history of these 5 species of primates that is consistent with the theory of evolution.  
\begin{table}
	\centering	
	\label{descriptive_stats}
	\begin{tabular}{lcccc}
		\toprule
		& Minimum & Mean   & Std. Dev. & Maximum \\
		\midrule
		Randomized data & -53484  & -51737 & 80.03   & -51675  \\
		\bottomrule
	\end{tabular}
	\caption{Descriptive statistics for the estimated log-marginal likelihoods from the randomized datasets.}
\end{table}  

%%%%%%%%%%%%%%%%%%
%%% Conclusion %%%
%%%%%%%%%%%%%%%%%%

\section{Conclusion}

A brief introduction to randomization method has been given. Two toy examples have been studied to explain its logic.  Example 1  represented a case in which the treatment had an effect on the studied characteristic, while Example 2 presented a case when chance was acting alone.  Both examples aimed to set up the logic which is used in the analysis of a primate family dataset.

We analysed a real dataset of 5 species of primates under a Bayesian statistical approach and used randomization to detect if this contained nonrandom information.  The data were permuted to eliminate any kind of phylogenetic signal, and then the probability that these randomized data came from the tree model was calculated (marginal likelihood).  This procedure was repeated several times, generating a distribution for the estimates.  The probability for the original dataset was much higher than the maximum value of the same value of the randomized data.  We would not expect such a probability if there was no tree signal.  Therefore, we concluded that chance was not acting alone and these species have a treelike relationship.  The presence of a hierarchical structure provides evidence for descent from common ancestry.  

The results given here are consistent with the theory of evolution and are added to the huge amount of evidence which supports it.  For instance, 28 morphological datasets were analyzed and are in favor of the treelike models \citep{Archie:1989}; in addition sequence data for 5 proteins from 11 species contains similar phylogenetic information \citep{Penny:Foulds:Hendy:1982}.  In this line, we have shown that Bayesian inference provides the means to detect this phylogenetic signal through the marginal likelihood.  In practice, it is unusual to find data that completely lack hierarchical structure \citep{Baum:2012} and the data analyzed here were not the exception. \\ % page 271 of Tree Thinking

All the analysis and plots have been produced in \textsf{R}-project \citep{R_project}.

%%%%%%%%%%%%%%%%
%%% Appendix %%%
%%%%%%%%%%%%%%%%

\section*{Appendix 1}
\addcontentsline{toc}{section}{Appendix}

%%% 1 %%%

\label{app1}
We analyze the dataset assuming a GTR+$\Gamma_4$ model and consider the following prior distributions on the parameters involved in the analysis:

\begin{itemize}
	\item Branch lengths: $t_i|\mu \sim$ Exp$(1/\mu)$, for $i =1,\dots,8,$ with $\mu \sim$ Inverse-Gamma(3,0.2).\\
	\item Relative rates: $q_i|\phi \sim$ Exp$(\phi)$, for $i =1,\dots,5,$ with $\phi\sim$ Exp(1).\\
	\item Base frequencies: $\pi\sim$ Dirichlet(1,1,1,1).\\
	\item Gamma shape parameter: $\lambda \sim$ Gamma(0.5,1).\\
\end{itemize}

For more information about the parameters involved in the phylogenetic analysis, see \cite{Yang:2014}.

%%% 2 %%%
\section*{Appendix 2}
\label{app2}
Nested sampling \citep{Skilling:2006} is a Bayesian algorithm to estimate mainly the marginal likelihood.  It requires a tunning parameter called \textit{active points}.  The precision of the estimate depends on the number of active points.  The higher it is, the more accurate the estimate and the higher the computational cost are.  

To estimate the observed marginal likelihood, we use 100 active points.  This yields a standard deviation of 0.73 of the log-marginal likelihood estimate.  For the 1000 randomized datasets, we use 5 active points in order to get a quick picture of their log-marginal likelihood distribution.

%%%%%%%%%%%%%%%%%%%%%%%%%%%%%%%%
%%%%%%%%%% References %%%%%%%%%%
%%%%%%%%%%%%%%%%%%%%%%%%%%%%%%%%

%\section{References}
%\nocite{*}
%\renewcommand{\section}[2]{}%
\bibliographystyle{sysbio}
\bibliography{Randomization_phylo}

\end{document}